# Strontium Iodide Radiation Instrument (SIRI) – Early On-Orbit Results

Lee J. Mitchell[a], Bernard F. Phlips[a], J. Eric Grove[a], Theodore Finne[a], Mary Johnson-Rambert[a],
W. Neil, Johnson[b]

[a] *United States Naval Research Laboratory, 4555 Overlook Ave. S.W., Washington, DC 20375*
[b] *Praxis Inc., 251 18th Street South, Suite 610, Arlington, VA 22202*

*Abstract—* The Strontium Iodide Radiation Instrument (SIRI) is a single detector, gamma-ray spectrometer designed to space-qualify the new scintillation detector material europium-doped strontium iodide ($SrI_2$:Eu) and new silicon photomultiplier (SiPM) technology. SIRI covers the energy range from .04 - 8 MeV and was launched into 600 km sun-synchronous orbit on Dec 3, 2018 onboard STPSat-5 with a one-year mission to investigate the detector's response to on-orbit background radiation. The detector has an active volume of 11.6 cm$^3$ and a photo fraction efficiency of 50% at 662 keV for gamma-rays parallel to the long axis of the crystal. Its spectroscopic resolution of 4.3% was measured by the full-width-half-maximum of the characteristic Cs-137 gamma-ray line at 662 keV. Measured background rates external to the trapped particle regions are 40-50 counts per second for energies greater than 40 keV and are largely the result of short- and long-term activation products generated by transits of the South Atlantic Anomaly (SAA) and the continual cosmic-ray bombardment. Rate maps determined from energy cuts of the collected spectral data show the expected contributions from the various trapped particle regions. Early spectra acquired by the instrument show the presence of at least 10 characteristic gamma-ray lines and a beta continuum generated by activation products within the detector and surrounding materials. As of April 2019, the instrument has acquired over 1000 hours of data and is expected to continue operations until the space vehicle is decommissioned in Dec. 2019. Results indicate $SrI_2$:Eu provides a feasible alternative to traditional sodium iodide and cesium iodide scintillators, especially for missions where a factor-of-two improvement in energy resolution would represent a significant difference in scientific return. To the best of our knowledge, SIRI is the first on-orbit use of $SrI_2$:Eu scintillator with SiPM readouts.

*Index Terms—* **astrophysics, background, gamma-ray, nuclear activation, radiation detector, scintillator, silicon photomultiplier, South Atlantic Anomaly, strontium iodide, Van Allen radiation belts**

## I. Introduction

SPACE qualification of new detector technologies for astrophysics or Department of Defense (DoD) applications are often a major challenge. The harsh space-radiation environment can have unanticipated effects on materials and devices originally developed for terrestrial applications. Before any new technology can be used in large-scale missions, it is typically subjected to extensive simulations and rigorous laboratory testing to reduce risk. However, it can be difficult to replicate the exact conditions encountered in space, and new missions often forego new technologies for heritage or proven components to reduce costs associated with drawing down that risk.

A number of new gamma-ray scintillation materials look promising and have been proposed for space missions [1] [2] [3] [4], while silicon photomultiplier (SiPM) readout technologies are also quickly replacing traditional photomultiplier tubes (PMTs) in instrument concepts [5] [6] [7]. The goal of the Strontium Iodide Radiation Instrument (SIRI) mission is to study the performance of new SiPM technology and a new scintillation material, europium-doped strontium iodide ($SrI_2$:Eu), for space-based gamma-ray spectrometry. $SrI_2$:Eu with SiPM readout offers improved energy resolution, lower power consumption, and reduced size over traditional PMT-based sodium iodide systems.

The Space Test Program (STP) operated by the DoD was designed to take newly developed, commercially available technologies that are promising for space applications and rapidly space-qualify them. STP missions typically last three years from space vehicle manifest to end-of-life, including two years for development and assembly of the instrument and satellite and one year of mission operations. SIRI is onboard STPSat-5, which has a sun-synchronous orbit at 600 km and hosts five payloads with a wide array of objectives [8]. The satellite was launched from Vandenberg Air Force Base, on December 3, 2018 as part of the Spaceflight SSO-A: SmallSat Express onboard a Space X's Dragon 9 [9] [10]. SIRI operations began after the extended commissioning period of the spacecraft in late January 2019. SIRI has been on and operating when the spacecraft is in nominal operations mode.

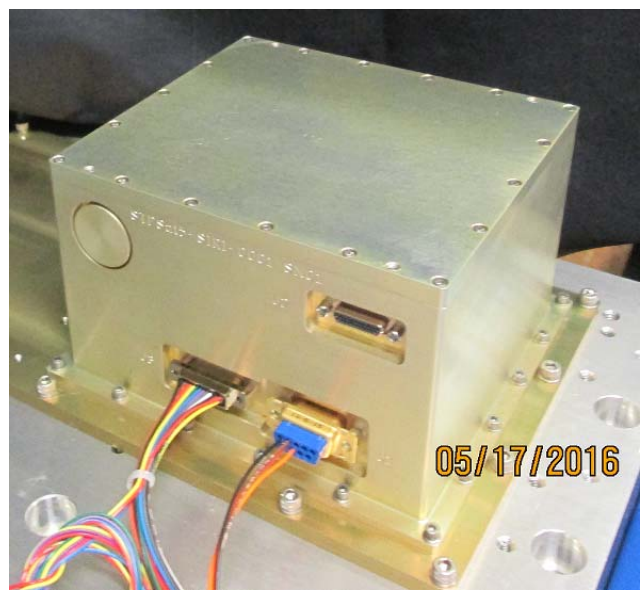

Fig. 1. Assembled SIRI instrument undergoing functional testing prior to spacecraft integration.

## II. INSTRUMENT DESCRIPTION

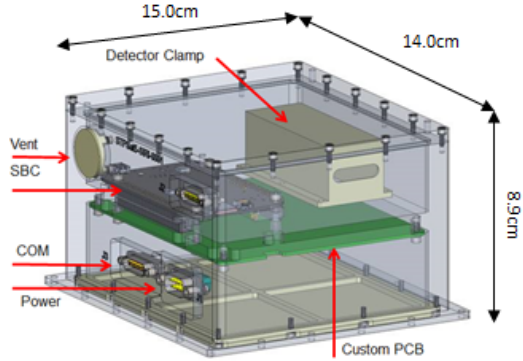

*Fig. 2. CAD rendering of SIRI showing the overall instrument design and the locations of major components.*

A brief system overview is provided here; details on the instrument hardware and software may be found in an earlier publication [11]. The SIRI instrument, shown in Figure 1, is a simple gamma-ray spectrometer built almost entirely of commercial-off-the-self (COTS) products. This allowed for rapid prototyping and construction of the instrument to meet the two-year time frame from start of design to delivery. Figure 2 shows the major components of SIRI in a CAD rendering; consisting of the detector, a multichannel analyzer (MCA), and a single board computer (SBC). A Kromek K102 MCA provides digitized events of the detector signal to the SBC, a BeagleBone Black. The SBC handles event timing, histogramming, and housekeeping, and communicates the data to the spacecraft over a RS422 port for eventual transmission to ground based receivers. The dimensions of the instrument are 8.9 cm (H) x 14.0 cm (D) x 15 cm (L), and the total mass is 1.6 kg. The nominal power is 1.5 W.

The SIRI detector assembly is composed of a 17mm x17mm x 40mm $SrI_2$:Eu crystal hermetically sealed in a 1mm aluminum housing and viewed via a 1/8" quartz window coupled to the crystal with a Sylgard pad. The assembly was manufactured by Radiation Monitoring Devices Inc. (RMD). A 2 x 2 array of SensL 6-mm J-series SiPMs are bonded to the quartz window with optically transparent Dow Corning 93500 silicone elastomer. In 2015, this was the largest commercially-available $SrI_2$:Eu crystal that could be procured and had an energy resolution of 4.3% at 662 keV. The detector is mounted to a mid-deck of the 6 mm thick aluminum instrument enclosure. The J-series SiPMs convert the scintillation light generated in the crystal to analog signals processed by the MCA. An AiT HV80 power supply provides a 29 V bias to the SiPM, and can be used to tailor the gain. The energy range of the spectrometer is .040 to 8 MeV – 4.0 MeV at the nominal voltage setting – and the detector has a photopeak efficiency of approximately 50% at 662 keV for gamma rays incident along the long axis of the detector. An overflow bin captures all events exceeding the dynamic range set by the commanded bias voltage. For comparison purposes, the light output of commercially available $SrI_2$:Eu is greater than 80,000 photons/MeV [12], a factor of two greater commercially available thallium-doped sodium iodide NaI(Tl) which is 38,000 photons/MeV [13].

Custom pre-amp and shaping electronics between the SiPM and the MCA were required to optimize the signal to meet the input requirements of the MCA. A Kromek K102, a single 4K multichannel analyzer (MCA) is used to convert the analog signals from the SiPM readout to time-tagged events. These digitized events are read out from the MCA by the onboard SBC. The SBC histograms these events based on a configurable integration time, and both the time-tagged events and histograms may be telemetered to ground along with housekeeping data. At low count rates, outside the trapped particle regions, all time-tagged events and histograms are packetized and transmitted. However, when the count rate exceeds a configurable threshold (nominally 2000 cps) for a configurable duration (nominally 10 s), SIRI autonomously stops forwarding time-tagged event data until the rate drops below that threshold for a configurable time (again, nominally 10 s). During this time, SIRI forwards only histograms. This prevents SIRI from exceeding the 115 kbps limit to the spacecraft as well as the nominal 82 (maximum 200) MB daily data volume. This software-controlled feature keeps SIRI from exceeding its allocated bandwidth during transits of the Van Allen belts and still allows the system to potentially capture event data during intense short gamma-ray transients such as Gamma Ray Bursts (GRBs).

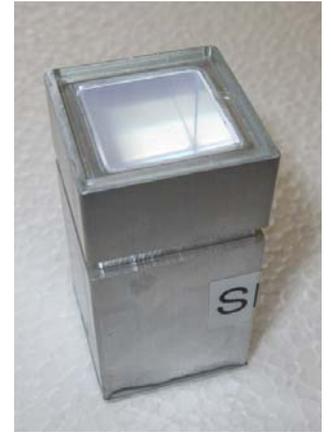

*Fig. 3. SIRI detector housed in an aluminum hermetically sealed container with a quarts optical window on one end. Not shown is the SensL J-series 2x2 6mm SiPM array bonded to the optical window and used as the readout.*

## III. EARLY ON-ORBIT RESULTS

### A. General Environmental Parameters

| Perigee | 575.9 km |
|---|---|
| Apogee | 597.9 km |
| Inclination | 97.7 deg |
| Period | 96.3 min |
| Semi-major axis | 6957 km |

*Table. 1. STPSat-5 orbital parameters based on tracking data provided by the US Air Force (USAF).*

The instrument boots and begins data collections reliably and without incident on orbit. To date, there have been no issues in resuming science ops after spacecraft transitions from safehold to nominal and no evidence of the single board computer or DAQ (BeagleBone) having issues over the first 6 months. At commissioning, on-orbit power consumption, gain, and noise were consistent with ground observations.

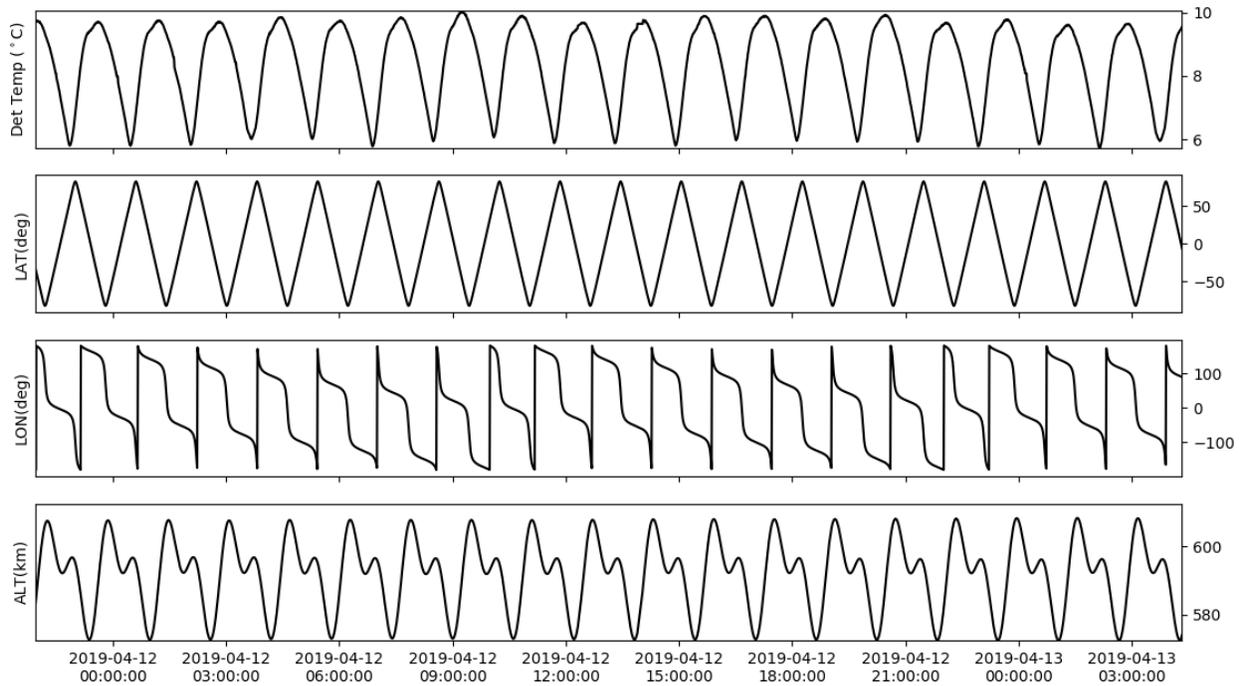

*Fig. 4. SIRI on-orbit parameters over a ~24hr period. Shown are values for temperature, latitude, longitude and altitude as measured by onboard sensors.*

Table 1 lists the orbital parameters based on tracking data from the USAF. The attitude of the spacecraft is such that the normal to the payload deck maintains a ram direction, a requirement unrelated to the SIRI experiment. Figure 4 shows the typical variation in temperature, latitude, longitude and altitude over a 24-hour period as measured by onboard sensors. The detector temperature is measured by a thermistor bonded to the detector clamp, shown in Figure 2. The measured value oscillates between 6ºC and 9.5ºC each orbit and appears to track with latitude as can be seen in Figure 4. This profile is expected, as the solar flux incident on the spacecraft is cyclic. These data were taken in southern summer, and the instrument is warmer at negative latitudes, as expected.

The light output of the strontium iodide crystal is known to be a function of temperature and is the predominant factor in gain change. According to the manufacturer specifications, the bias power supply is stable to 2 ppm/ºC and is not expected to affect the gain. SIRI has no on-board gain stabilization. Instead, a calibration profile was developed prior to launch from the thermal vacuum test data and is used to calibrate the instrument in post processing of the data on the ground. The 511 keV annihilation line is also used to verify the calibration on orbit. Figure 5 shows the centroid of the 511 keV peak, plotted as a function of temperature. Spectra were generated for four temperature ranges, and each 511 keV peak was fitted using a Gaussian to determine the centroid. Average temperature values over the individual ranges where used for abscissa values.

Figure 6 shows the centroid of the 511 keV line as function of geolocation and represents an average over a three month time period. To generate the map, spectra were summed for each pixel and a Gaussiam fit on the 511 keV line was used to determine the centroid. In the northern hemisphere the spacecraft is cooler, and in the southern hemisphere it is warmer. This leads to less than 4% variation in gain of the 511 keV peak over a given orbit. In terms of energy, this corresponds to a shift range of 15-18 keV. Areas that could not be fitted due to high background are indicated by dark blue, such as the South Atlantic Anomaly (SAA) and the polar belts.

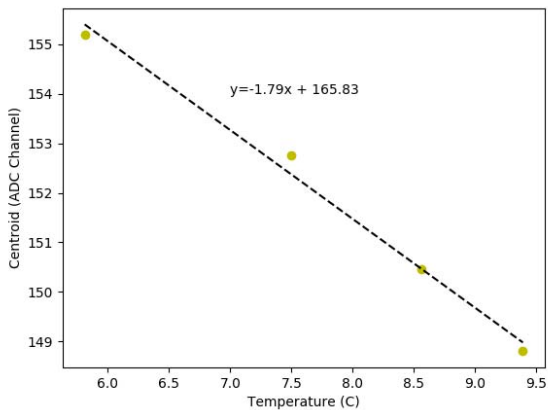

*Fig. 5. Centroid of the 511 keV line as a function of temperature. Plot shows the variation in detector gain which is corrected in post processing on the ground.*

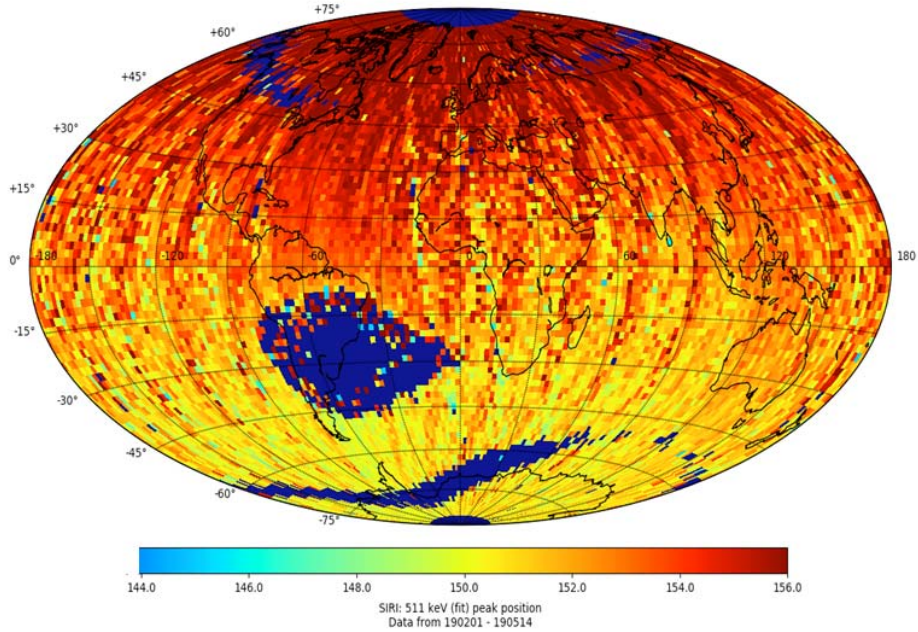

*Fig. 6. Map of the 511 keV centroid channel with respect to geolocation, showing the effect of temperature on gain. In regions such as the SAA and polar belts, a fit could not be performed and are indicated by a dark blue.*

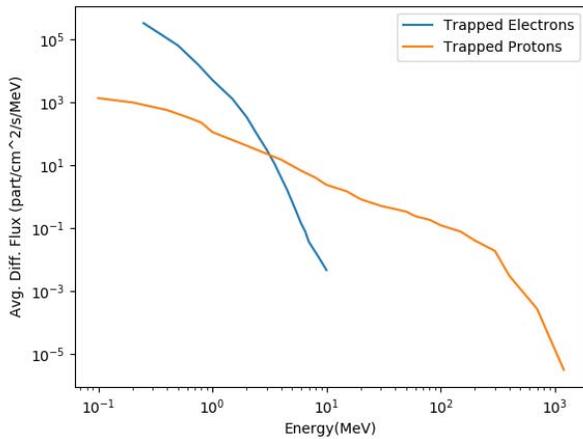

*Fig. 7. AE9 and AP9 predicted average differential fluxes for both trapped electrons and protons. Latitude, longitude, and altitude data from the month of April 2019 was used as the input ephemeris file for the model.*

### B. On-orbit Background

SIRI's sun-synchronous orbit results in the instrument traversing the South Atlantic Anomaly (SAA) and the polar radiation belts. These regions of trapped particles, composed primarily of protons and electrons, are the largest contributors to background radiation observed by SIRI and are part of the Van Allen belts [14]. The SAA is a region of the inner Van Allen belts that extends down to lower altitudes. In SIRI's orbit the SAA is most intense over South America and gradually declines in intensity as it extends out over the Atlantic towards Africa. The background in low-earth orbit (LEO) is known to be dependent on altitude and inclination [15], with solar activity providing a temporal component. Outside of these trapped particle regions, residual activation and galactic cosmic rays are the major components to the background, with the diffuse galactic gamma-ray background and secondary gamma rays from cosmic-ray interactions in the Earth's atmosphere playing a minor role.

Figure 7 shows the differential fluxes for both trapped electrons and protons as predicted by the AE9 and AP9 near-Earth radiation models [16] for the ephemeris of STPSat-5 for April 2019. The fluxes represent an average for that month, and are plotted over the range of 0.25 to 10 MeV and 0.1 to 1200 MeV for trapped electrons and protons, respectively. Changes in the trapped particle fluxes along the orbital path leads variablity in the SIRI on-orbit background.

Figure 8 shows an intensity map of the gross count rate for energy depositions greater than 40 keV measured by the SIRI detector and was generated from 30 second spectra periodically produced onboard the instrument. Data was collected for from 60 days of operations since the instrument's commissioning. Areas where no data was collected are white (detector bias is set to 0 during the most intense portions of the SAA). Four zones marked as A, B, C and D were used to generate the spectra shown in Figure 9 and are representative of the wide variation in background experienced by SIRI.

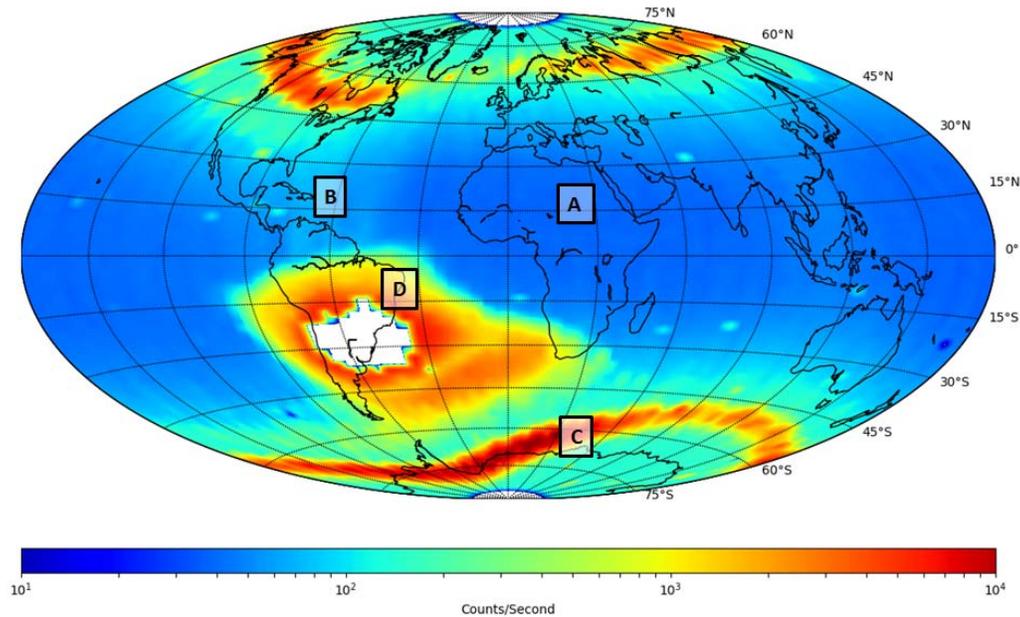

Fig. 8. Gross gamma-ray count rate showing the elevated background as the instrument transitions through the various trapped particle regions. The four zones A, B, C and D were used generate the spectra shown in Figure 5. No data is indicated by the white areas of the plot.

Zones A and B are representative of the background spectra external to the trapped particle regions with the latter displaying a slightly higher intensity due to activation of the instrument and spacecraft after transits of the SAA. Traditional LEO-based gamma-ray instruments operate in Zone A and B environments. Preferred orbits have a low inclination (as close to equatorial as possible) and the instrument is typically powered down through the SAA. Zones A and B are typical of the minimum and maximum backgrounds one would expect while observing astrophysical phenomena with a comparable $SrI_2$ detector.

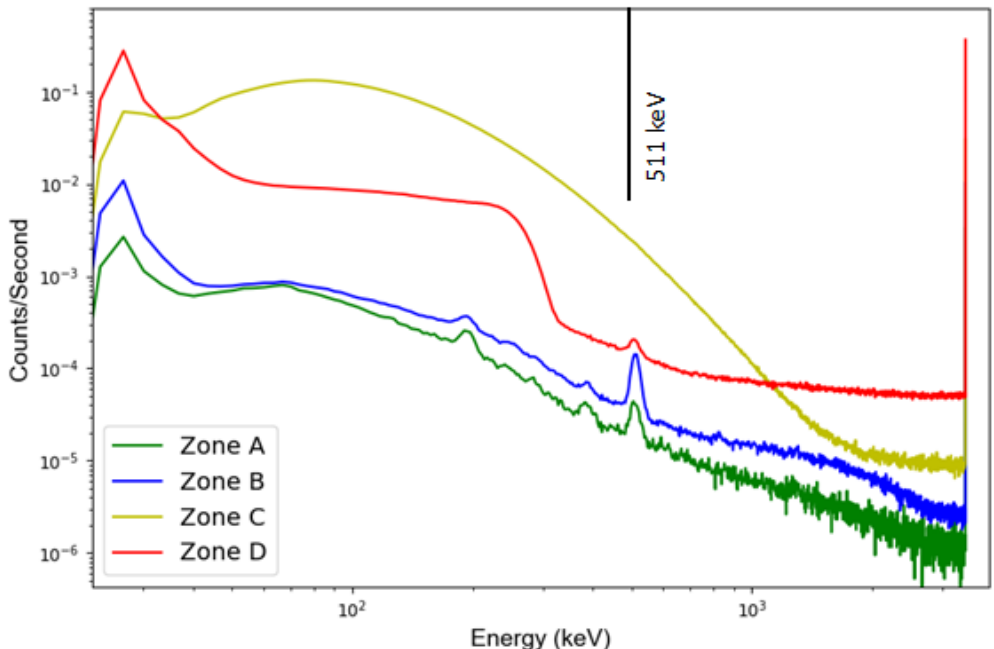

Fig. 9. Summed spectra generated from zones indicated in Figure 8, showing the wide variation in the on-orbit background measured by SIRI.

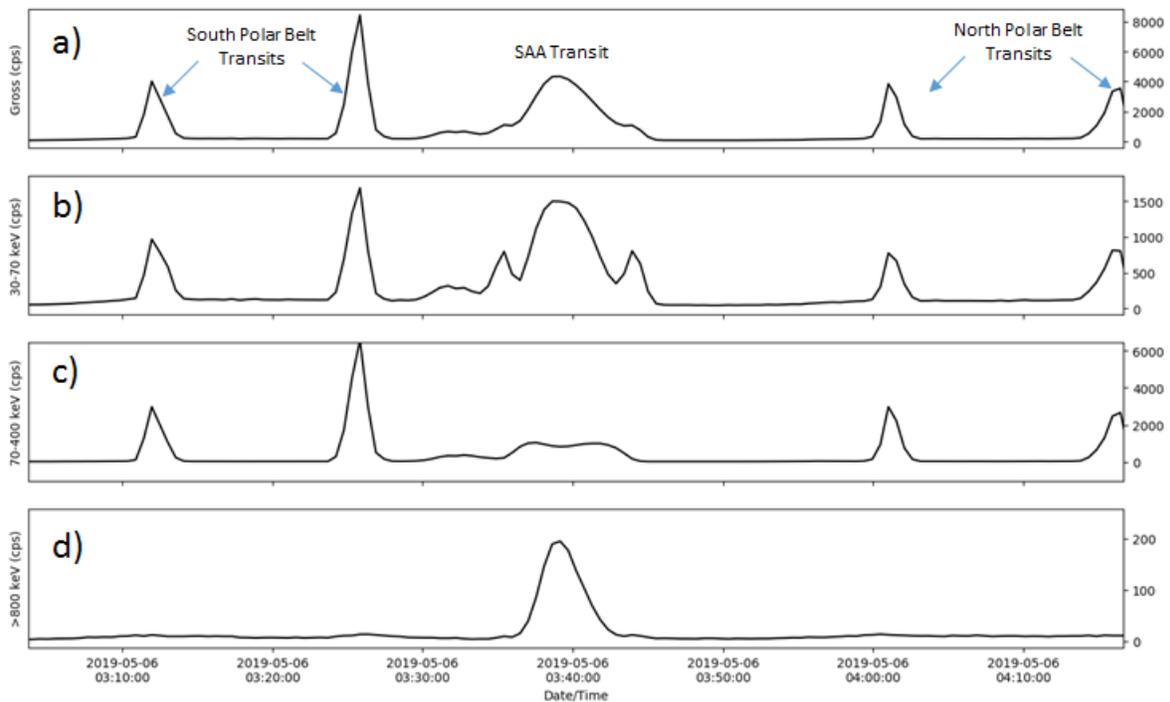

*Fig. 10. Rate plots in counts per second generated from the 30 second spectra over a typical orbit. The gross rate(a) and rates corresponding to spectral energy cuts of 30-70 keV (b), 70-400 keV(c) and >800 keV(d) are plotted.*

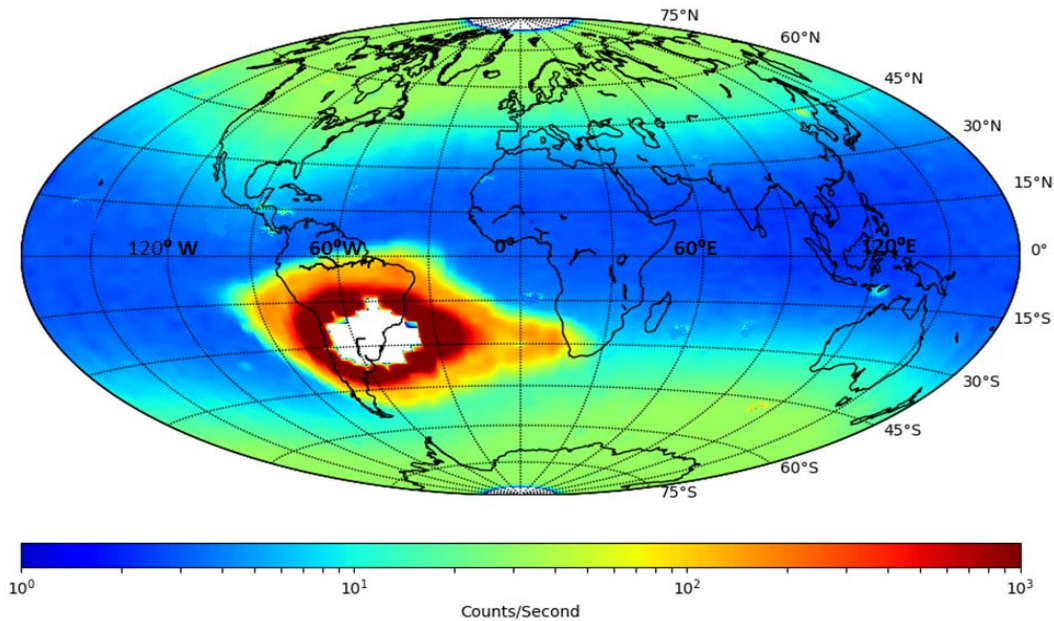

*Fig. 11. Intensity rate map of the overflow bin showing detector events greater than 4 MeV. Cut emphasizes energetic protons.*

The spectrum in Zone C corresponds to the polar belts, and the shape indicates a Bremsstrahlung spectrum generated by a distribution of energetic trapped electrons interacting with the surrounding materials of the instrument and spacecraft. The trapped electron regions can be seen in Figure 8, between 45° and 75° degrees latitude for the northern hemisphere and between -45° and -75° for the southern hemisphere. These regions of intense radiation are where the outer Van Allen belts extend down to lower altitudes, and like the SAA are known to fluctuate in intensity based on solar weather [17]

[18]. While the total counts rates are high here (exceeding 20K cps), SIRI can safely operate in these regions without risk to the instrument, forwarding only histograms.

Zone D shows a spectrum within high-rate portions of the SAA where SIRI is collecting data. The spectral shape is indicative of a beta continuum, likely emitted by activation products generated by (p,xn) nuclear reactions induced by trapped protons. The count rates in this region of 1 to 6K cps are substantially less than those observed in the trapped electron regions of Zone C. Above energy depositions of ~1 MeV the continuum is relatively constant and many events have energies in the overflow bin of the spectrum. This is consistent with the signal expected from energetic trapped protons depositing a large amount of energy in the crystal.

While the SAA is composed of both protons and electrons, the protons are the greatest concern for SIRI's detector as the electrons in this region are low energy and are strongly attenuated by the 7 mm of aluminum surrounding the $SrI_2$ crystal. Simulations with Stopping and Range of Ions in Matter (SRIM) show that protons above 40 MeV penetrate the surrounding aluminum [19]. While these protons represent only a small portion of the overall distribution as seen in Figure 7, these highly energetic particles deposit a large amount of energy and a proportional amount of scintillation light is generated when they interact with the detector crystal. This can lead to an increase in the current demand on the bias power supply from the SiPM. To combat this problem, SIRI is commanded to turn off the detector bias supply during transits of the more extreme portions of the SAA. A polygon boundary is used to define this region and absolute time commands, generated from predicted ephemerides, are used to turn the detector off and on upon entry and exit of the SAA.

Figure 12 plots from the rate of the the overflow bin (>4MeV) from the periodically produced 30s spectra as a function of the McIlwain L-parameter. The inner and outer van Allen belts are roughly between L=1-3 and L=4-6, respectively. At L-values greater than 4, the cosmic-ray flux is no longer modulated by the Earth's magnetic field. The turnover corresponds to a vertical cut rigidity of 900 MeV which in agreement with the current solar-cycle modulation of galactic cosmic-rays. Bremsstrahlung from energetic electrons in the outer Van Allen are excluded due to the 4 MeV cut.

Figure 10 shows SIRI rate plots with various energy cuts over a "typical" orbit and were generated from the periodically produced onboard 30 second spectra. The top plot shows the gross count rate with 4 transits of the polar belts and 1 transit of the SAA. Astrophysical observations would be most sensitive during those portions of the orbit when the background rate is low. External to the trapped particle regions, an average minimum total rate of ~80 cps is observed. The energy between 30-70 keV represents about 50 percent of the total background measured. From the spectra in Figure 9 it is easy to see how an energy cut between 70-400 keV emphasizes the polar belt transits over the SAA as it cuts out the energetic protons, and an energy cut above 800 keV excludes the bremsstrahlung and emphasizes the SAA and cosmic-ray contributions.

Figure 11 shows an intensity map of the count rate of the overflow bin. At energies above 3.5 MeV the majority of events detected by SIRI are the trapped protons in the SAA.

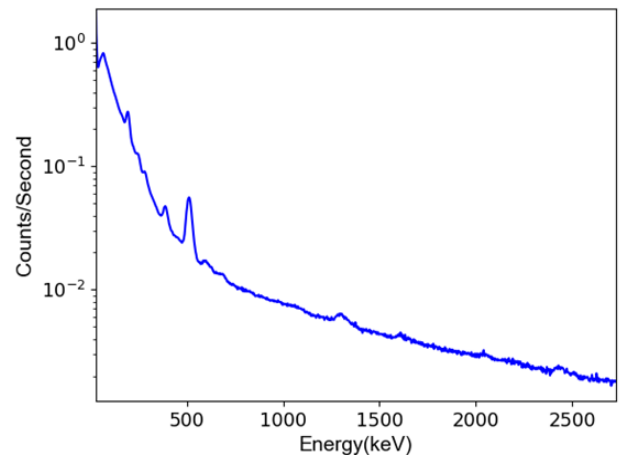

Fig. 13. Summed spectrum generated by spectra with less than 60 cps. The 511 keV is the predominant line, however at least 10 other lines due to likely activation products are visible.

However, the rate is also observed to increase when going from the equator to the poles. Cosmic-ray rates are known to increase in intensity when going from the equator to the Polar Regions, a direct result of the Earth's magnetic field [20]. Note the elevated count rates in the polar belts due to Bremsstrahlung radiation have completely disappeared when comparing Figure 11 to Figure 8. With a dead time of <12 μs per event, pileup is not expected to be a significant issue at these rates

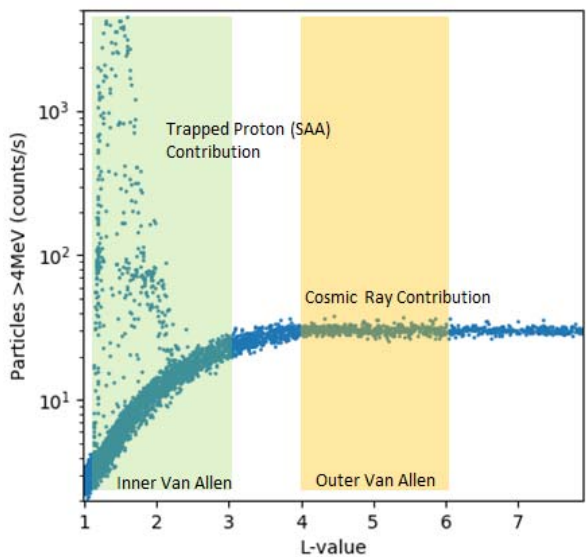

Fig. 12. Plot shows the counts/sec of the overflow bin (>4 MeV) of the periodically generated 30s spectra versus the McIlwain L- value..

*C. Spectroscopy*

The primary benefit of SrI$_2$:Eu relative to previous scintillators (e.g. NaI:Tl) is its improved energy resolution. Figure 13 shows a spectrum of the on-orbit background summed from portions of the orbit with low particle background (i.e. count rates less than 60 cps). Selectively choosing spectra from low background regions emphasizes the longer-lived activation products that contribute to this background. These activation products are mostly generated when the spacecraft travels through the SAA, but a smaller component produced by the continued bombardment of cosmic rays is also present. The 511 keV annihilation line is the dominant characteristic gamma ray in the spectrum and is associated with beta decays of activation products in the detector and surrounding materials [21]. At least 10 other lines are visible in the spectrum. Activation products from (p,xn) reactions with iodine are commonly observed in NaI(Tl) instruments, and SIRI is expected to see similar activation products. As an example, the presence of characteristic gammas of 57.6 and 202.9 keV in Fermi Gamma-ray Burst Monitor data demonstrates that $^{127}$I is produced [22]. Detailed analysis of the activation products generated in SIRI will be the subject of a future paper.

Figure 14 shows an intensity map of the 511 keV annihilation line plotted as a function of geolocation. Spectral data for the first 3 months of operations were summed for each geospatial pixel. The 511 keV line in each summed spectrum was fitted with a Gaussian to determine the intensity. The background continuum was approximated by fitting a first-order polynomial to the continuum surrounding the peak. Significant statistical noise is visible in the plot; however, this will improve as SIRI collects more data. The regions north and south of the SAA show high 511keV line intensity. With the spacecraft in a sun-synchronous orbit, taking a path generally north-south through the SAA, this intense 511 keV line is thus the signature of activation of SIRI and the components of STPSat-5 by the intense flux of trapped protons in the SAA, likely β+ decays from the products of (p,xn) reactions. The spectrum from Zone B (Figure 9) shows a prominent 511 keV line from this activation.

IV. CONCLUSION

As of April 2019, SIRI has accumulated over 1000 hours of on-orbit background gamma-ray data. The system continues to perform as expected. Future work will consist of analyzing long-term trends in parameters such as detector energy-resolution and the operational characteristics of the SiPM. A detailed analysis of the activation products that generate the characteristic gamma rays seen in Figure 13 will be compared to Monte Carlo simulations in an effort to identify isotopes and their reactions. AE9 and AP9 trapped particle models will be compared to data collected to by the SIRI instrument as it transits the SAA and polar belts to better understand the trapped particle distribution creating the elevated background in the respective regions.

ACKNOWLEDGEMENTS

This work was supported by the Chief of Naval Research.

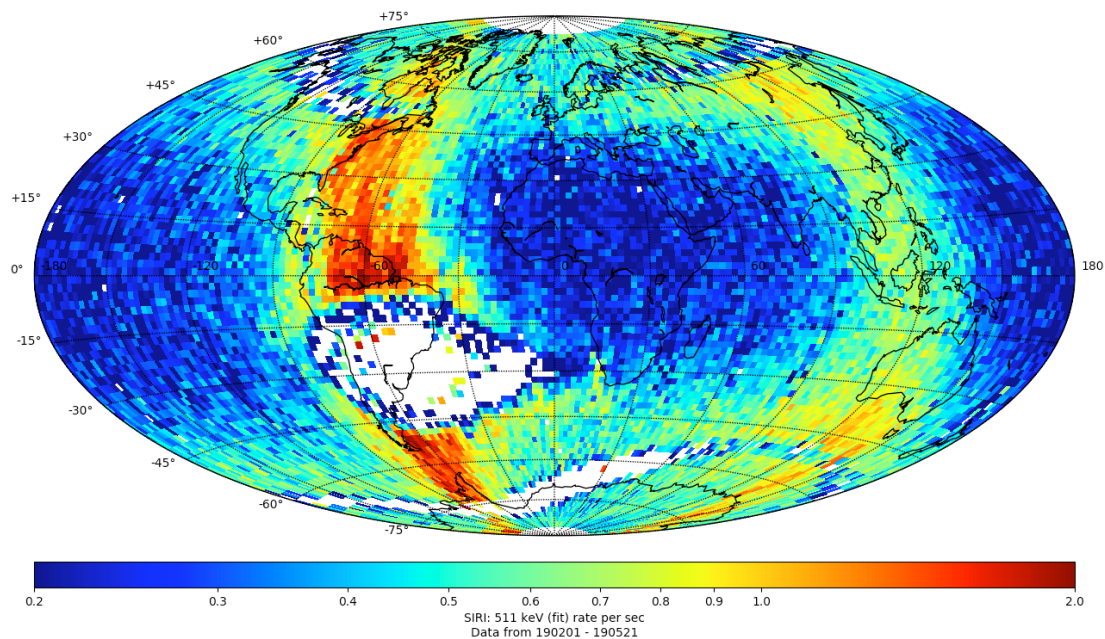

*Fig. 14. Intensity plot of the 511 keV line as a function of geographic location. Spectra were summed for each pixel and a Gaussian fit centered on the 511 keV line was used to measure the line intensity. Areas where it was not possible to fit the 511 line are indicated by white.*

REFERENCES

[1] P. F. Bloser, M. L. Cherry, G. L. Case, J. Greiner, G. Kanbach, R. M. Kippen, M. L. McConnel, R. S. Miller, J. M. Ryan, K. Shah, C. J. Stapels, E. V. Loef and M. Wallace, "Advanced Scintillators and Readout Devices for High-Energy Astronomy," National Academies, Washington, 2009.

[2] M. McConnell, P. Bloser, L. J. and J. Ryan, "Applications for New Scintillator Technologies in Gamma Ray Astronomy," *Journal of Physics,* vol. 763, no. 1, 2016.

[3] E.-J. Buis, M. Beijersbergen, S. Kraft, A. Owens, F. Quarati, S. Brandenburg and R. Ostendorf, "New scintillators for focal plane detectors in gamma-ray missions," *Experimental Astronomy,* vol. 20, pp. 333-339, 2005.

[4] R. S. Perea, A. M. Parsons, M. Groza, T. E. Peterson, D. Caudel, K. G. Stassun, S. Nowickib and A. Burger, "Scintillation properties of SrI2(Eu2+) for high energy astrophysical detectors: Nonproportionality as a function of temperature and at high gamma-ray energies," arXiv, Ithica, 2014.

[5] K. Lacombe, B. Houret, J. Knödlseder, J. Gimenez, J.-F. Olive and P. Ramon, "Characterization of silicon photomultipliers for new high-energy space telescopes," *Nuclear Instruments and Methods in Physics Research Section A,* vol. 912, pp. 144-148, 2018.

[6] M. Teshima, B. Dolgoshein, R. Mirzoyan, E. Popova and J. Nincovic, "SiPM Development for Astroparticle Physics Applications," arXiv:0709.1808 [astro-ph], Ithica, 2007.

[7] R. Wagner, K. Byrum, M. Sanchez, A. V. Vaniachine, O. Siegmund, N. Otte, E. Ramberg, J. Hall and J. Buckley, "The Next Generation of Photo-Detector for Particle Astrophysics," U.S. Department of Energy, Oak Ridge, 2009.

[8] Sierra Nevada Corporation, "New Sierra Nevada Corporation Satellite to Launch," SNC, 17 Nov. 2018. [Online]. Available: https://www.sncorp.com/press-releases/snc-stpsat-5-set-to-launch/. [Accessed 31 Mar. 2019].

[9] S. Clark, "SpaceX launches swarm of satellites, flies rocket for third time," 3 December 2018. [Online]. Available: https://spaceflightnow.com/2018/12/03/spacex-launches-swarm-of-satellites-re-flies-rocket-for-third-time/.

[10] SpaceFlight, "Instroducing SSO-A: The SmallSat Express," SpaceFligh, 2019. [Online]. Available: http://spaceflight.com/sso-a/. [Accessed 2019 February 2019].

[11] L. J. Mitchell, B. F. Phlips, R. S. Woolf, F. T. Theodore, N. J. Johnson and E. G. Jackson, "Strontium Iodide Radiation Instrumentation (SIRI)," in *Proceedings of SPIE*, San Diego, 2017.

[12] RMD Inc., "Gamma Scintillator Properties SrI2," [Online]. Available: https://cdn.dynasil.com//assets/Strontium-Iodide-SrI2-Gamma-Scintillator-Properties.pdf. [Accessed 31 03 2019].

[13] Saint Gobain, "NaI(Tl) and Polyscin® NaI(Tl) Sodium Iodide," [Online]. Available: https://www.crystals.saint-gobain.com/sites/imdf.crystals.com/files/documents/sodium-iodide-material-data-sheet_0.pdf. [Accessed 31 03 2019].

[14] J. A. van Allen, "Observation of high intensity radiation by satellites 1958 alpha and gamma.," *Journal of Jet Propulsion,* vol. 28, no. 9, p. 588–592, 1958.

[15] P. Cumani, M. Hernanz, J. Kiener, V. Tatischeff and A. Zoglauer, "Background for a gamma-ray satellite on a low-Earth," 19 Feb. 2019. [Online]. Available: https://arxiv.org/abs/1902.06944. [Accessed 31 Mar. 2019].

[16] G. P. Ginet, S. L. Huston, T. P. O'Brien and W. R. Johnston, "AE9, AP9 and SPM: New Models for Specifying the Trapped Energetic Particle and Space Plasma Environment," *Space Science Review,* vol. 179, pp. 579-615, 2013.

[17] M. K. Hudson, B. T. Kress, H.-R. Mueller, J. A. Zastrow and J. B. Blake, "Relationship of the Van Allen radiation belts to solar wind drivers," *Journal of Atmospheric and Solar-Terrestrial Physics,* vol. 70, no. 5, pp. 708-729, 2008.

[18] J. Domingos, D. Jault, M. A. Pais and M. Mandea, "The South Atlantic Anomaly throughout the solar cycle," *Earth and Planetary Science Letters,* vol. 473, pp. 154-163, 1 Sep. 2017.

[19] J. F. Ziegler, J. P. Biersack and M. D. Ziegler, SRIM - The Stopping and Range of Ions in Matter, Lulu, 2015.

[20] D. J. HOfmann and H. H. Sauer, "Magnetospheric Cosmic-ray Cutoff and Their Variations," *Space Science Reviews,* vol. 8, no. 5-6, pp. 750-803, 1968.

[21] C. Meegan, G. Lichti, P. N. Bhat, M. S. Briggs, V. Connaughton and et al., "The Fermi Gamma-ray Burst Monitor," *The Astrophysical Journal,* vol. 702, pp. 791-804, 2009.

[22] C. S. Dyer, J. I. Trombka, S. M. Seltzer and L. G. Evans, "Calculation of radioactivity induced in gamma-ray spectrometers during spaceflight," *Nuclear Instruments and Methods,* vol. 173, no. 3, pp. 585-601, 1980.